# Morphological Evolution of NMC Secondary Particles Through *in situ* electrochemical FIB/SEM experiment

*François Cadiou[5], Tuan-Tu Nguyen[5], Martin Bettge[2], Zeliang Su[5], Jonathan Ando[4], Vincent De Andrade[3], Dean Miller[1*], Arnaud Demortière[1,4,5,6*]*

[1]Electron Microscopy Center, & Engineering, Argonne National Laboratory, 9700 S. Cass Ave., Argonne, IL 60439, USA

[2]Chemical Sciences, Argonne National Laboratory, 9700 S. Cass Ave., Argonne, IL 60439, USA

[3]Advanced Photon Source, Argonne National Laboratory, 9700 S. Cass Ave., Argonne, IL 60439, USA

[4]Réseau sur le Stockage Electrochimique de l'Energie (RS2E), 80039 Amiens Cedex, France

[5]Laboratoire de Réactivité et Chimie des Solides (LRCS) CNRS, UMR7314, Univ. Picardie Jules Verne, 80039 Amiens Cedex, France

[6]ALISTORE-European Research Institute, CNRS FR 3104, Hub de l'Energie, Rue Baudelocque, 80039 Amiens Cedex, France.

Corresponding authors: dean.miller@tescan.com, arnaud.demortiere@cnrs.fr

**Abstract:** Microstructural evolution of NMC secondary particles during the battery operation drives the electrochemical performance and impacts the Li-ion battery lifetime. In this work, we develop an *in situ* methodology using the FIB/SEM instrument to cycle single secondary particles of NMC active materials while following the modifications of their 3D morphology. Two types of secondary particles, *i.e.* low and high gradient NMC, were studied alongside morphological investigations in both pristine state and different number of cycles. The quantification of initial inner porosity and cracking evolution upon electrochemical cycling reveals a clear divergence depending on the type of gradient particles. An unexpected enhancement of the discharge capacity is observed during the first cycles concurrently to the appearance of inner cracks. At the first stages, impedance spectroscopy shows a charge transfer resistance reduction that suggests a widening of the crack network connected to the surface, which leads to an increase of contact area between liquid electrolyte and NMC particle. 3D microstructure of individual secondary particles after *in situ* cycles were investigated using FIB/SEM and nano-XCT. The results suggest a strong impact of the initial porosity shape on the degradation rate.

Lithium-ion batteries are now widely used in electronic portable devices. They have also been studied in the last decades as suitable candidates for electric vehicles (EV) powering due to their high energy and power densities as well as long cycle life and low cost. Among various active materials (AM) for cathode electrode, NMC ($Li(Ni_xMn_yCo_{(1-x-y)})O_2$) are undergoing a lot of research for EV usage, regarding different Ni, Mn and Co proportions [1–12]. In lithium-ion positive electrodes, the NMC type AM comes in the form of cluster-like secondary particles aggregated from primary grains. The element proportions (Co/Mn) can be either constant inside the secondary particles[1,3–8] or designed as a composition gradient particle, in which the composition varies from core to shell structure [2,4,9–12] to mitigate transition metal dissolution in the electrolyte and to improve thermal stability. The secondary particles are building blocks for electrode architecture and their properties are very important as they impact the microstructure and the percolation network of composite electrodes[13a,13b]. The presence of defects inside and at the surface of secondary particles, such as crack, porosity, and grain boundary, is closely related to transformation through electrochemical cycling leading to evolution of mechanical and chemical heterogeneity.[1–4,6,8,14–25].

In the last decade, fundamental studies of electrochemical phenomena have been slowed down by a lack of effective *in situ* and *operando* experimental setups, which are able to clearly identify structural modifications inside electrode materials. Microstructural evolutions, crack, and porosity appearance, SEI formation at the electrode/electrolyte interface, and crystal phase transformations must be properly investigated to get a better insight into the influence of charge/discharge processes on reaction mechanisms implied in electrochemical storage. Improving our understanding of the microstructural changes and crack formation in Li-ion electrode materials during electrochemical cycling can provide new strategies to optimize electrode fabrication. Kim *et al.*[26] show how cracks propagate inside secondary particles along the grain boundary defined by the morphology and orientation of the primary particles. A core-shell structure, with elongated grains in the shell and randomly oriented grains in the core, can lead to a limitation of the crack propagation towards the surface during prolonged cycles. An interesting alternative is the single crystalline particles[27], in which the lack of grain boundaries does not allow the formation of cracks, improving the electrochemical and thermal properties. However, single-crystalline cathodes need in-depth investigations to improve their performance at higher Ni composition (>80%) and their long-term cycling. Mao *et al.*[28a] investigated the 3D morphological defect within the NMC secondary particles using nano-resolution 3D TXM and machine learning methodology. They showed that cracks are highly dependent on the initial state 3D morphology and particles having inner porosity seem to be more robust against crack formation than the solid ones during the electrochemical cycling conditions. The morphology engineering of secondary particles should be more considered mitigating the crack damage to improve the performance of cathode electrodes. Coupled with contrast techniques, such as Zernike phase contrast and XANES spectroscopy, Nano-XCT have been recently used to characterize the complex 3D architectures of Li-ion battery electrodes and secondary particles[28b,28c,28d].

To monitor microstructural evolution dynamically during electrochemical cycling, we developed a micro-scale battery set-up implemented within a FIB/SEM instrument [8,20,29–31]. One of the most recent works involving this technique was made by Amine *et al.* and is described in their paper[31]. The single particle of positive electrode oxide (NMC) materials with a size of 5-10 µm is attached to a metal probe (W) via a carbon bridge (GIC) to establish connection of good electronic conductivity. The micromanipulator allows moving the particle in the chamber and immersing it in an ionic liquid electrolyte (with low vapour pressure)[8], which is deposited on the counter electrode, *i.e.,* lithium metal. Electrochemical measurements are carried out using an ultra-low current instrument (biologic SP200) with a two-point connection configuration to the external probe tip and SEM metal holder. After immersing into liquid electrolyte and minimization of Li-metal/particle distance, the active material single particle is cycled in galvanostatic mode with a steady current around 1 nA, which matches a C-rate of about 1 based on the particle volume and the theoretical capacity[29].

The experimental setup geometry differs from a traditional battery geometry as there is a tip for the active material and a planar surface for the lithium content, hence there is a strong disproportion in the electric field. Experimental setup was thus optimised, mainly by reducing particle/Li-metal distance, to minimize the over potential and get a proper charge/discharge curve. We studied structural modifications in individual particles after each charge/discharge cycle by FIB slicing and SEM imaging, experimental details can be found in SI. We quantified the crack formation as a function of cycle number. Evolution of the discharge capacity was correlated with crack and porosity appearance inside the positive electrode particle. Impedance spectroscopy measurements suggested an overall decrease in charge transfer resistance for the particle that is linked to the initial stage of crack formation, which induces a discharge capacity enhancement. On the other hand, the 3D structural characterization of these materials is crucial to better understand the structural configuration and the evolution of the discharge capacity. For this purpose, 3D microstructural reconstruction from FIB tomography acquisition was used on single secondary particles[32]. Changes in structural parameters such as porosity, grain connectivity and crack propagation were extracted from the 3D reconstruction and related to electrochemical properties. Finally, 3D FIB/SEM data were compared with those obtained from the 3D TXM tomography[33], which was performed in beamline ID32 at APS synchrotron.

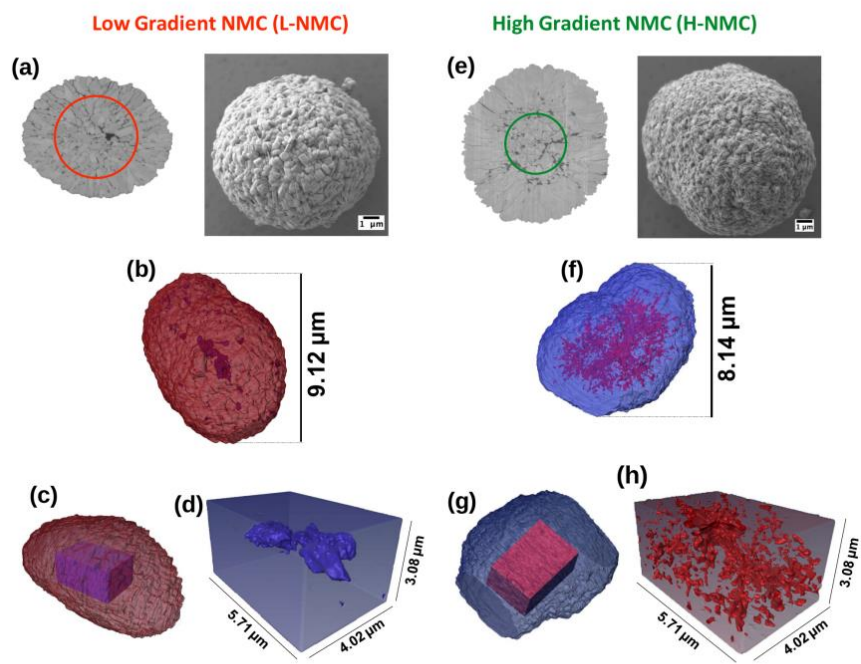

**Figure 1:** Visualizations for the two NMC particle types used. Left, Low Gradient particle (L-NMC) from (a) to (d), and right High Gradient particle (H-NMC) from (e) to (h). (a) and (e) display an internal (cutting slice) and external point of view in SEM microscopy, the coloured circles delimit the gradient areas (border between heart and shell). The other items are showing 3D transparent views of the particle highlighting the internal porosity and the region of interest for its analysis (boxes in (d) and (h)).

Two types of NMC secondary particle types were studied here. Although both are gradient-like secondary particles, one was a low gradient NMC particle (called L-NMC) and the other was a high gradient NMC particle (called H-NMC) as shown in Figure 1. L-NMC are composed of Li(Ni$_{0.68}$Co$_{0.18}$Mn$_{0.18}$)O$_2$ on average with a core Li(Ni$_{0.8}$Co$_{0.1}$Mn$_{0.1}$)O$_2$ and a Li(Ni$_{0.46}$Co$_{0.23}$Mn$_{0.31}$)O$_2$ outer shell while H-NMC are based on Li(Ni$_{0.75}$Co$_{0.1}$Mn$_{0.15}$)O$_2$ on average with a core Li(Ni$_{0.86}$Co$_{0.1}$Mn$_{0.04}$)O$_2$ and a Li(Ni$_{0.7}$Co$_{0.1}$Mn$_{0.2}$)O$_2$ outer shell. The secondary particles were typically about 8-9 µm in diameter and exhibited different internal porosity shapes depending on the type of particle. As shown in the 3D views and cross-sectional slices in Figure 1, the L-NMC particles exhibited bulkier porosity than the H-NMC ones where the porosity was more spider-web like. Both porosity types appeared approximately centred in the NMC secondary particle. This observation was made over several particles of each type and the porosity morphologies were found to be consistent across each particle formulation.

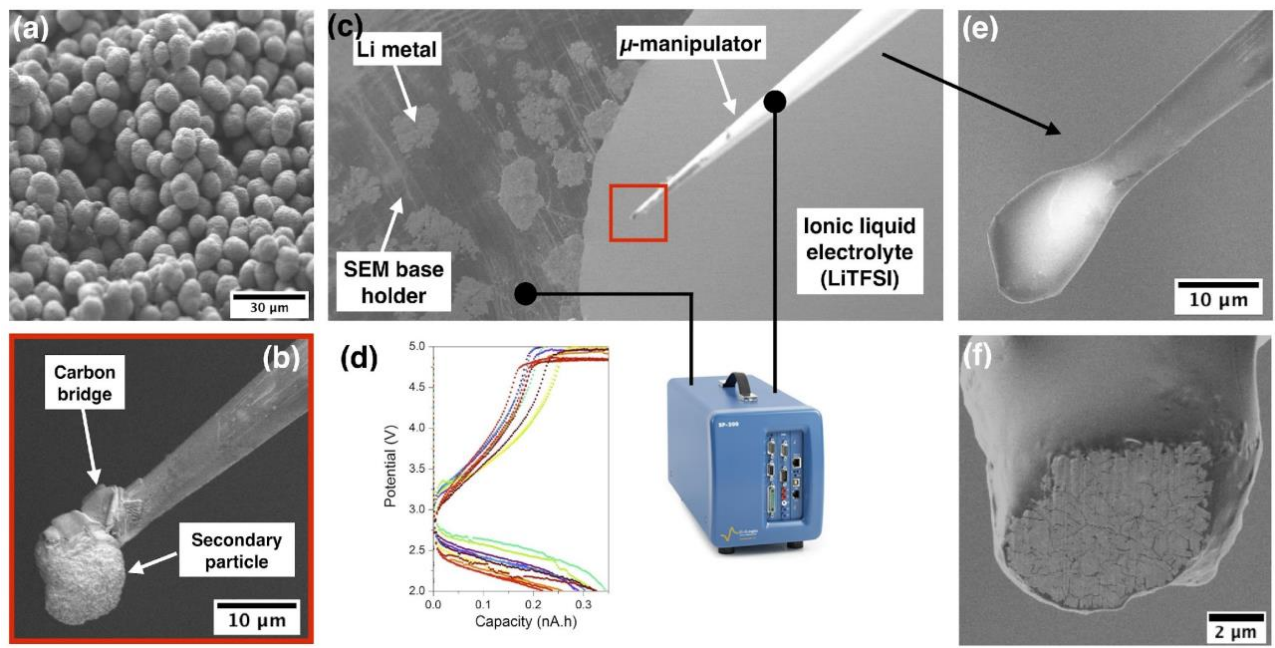

**Figure 2:** Experimental setup. (a) view of a NMC particle packing, (b) fixation of one single NMC particle on the micromanipulator tip (zoomed from red boxed area in (c)), (c) global view of the experimental area for cycling and connection points for cycling. (d) is displaying cycling curves obtained from the setup while (e) and (f) views of the NMC particle after cycling (e) external view, (f) sliced internal view).

The FIB/SEM experimental setup elements are described in Figure 2. Figure 2a shows an SEM view of an NMC secondary particle packing from which were extracted the single secondary particles to be cycled. These single secondary particles were attached to the SEM micromanipulator as illustrated in Figure 2b by forming a carbon bridge using a gas injection system (GIS). The attached single particles were immersed with the micromanipulator into a liquid electrolyte (LiTFSI ionic liquid) and stop to a position near (around 100 µm) to Li metal source (Figure 2c). The Li metal was previously electrochemically deposed on the holder in small domains as shown in Figure 2c. The low current potentiostat instrument is connected to the holder (anode) and the micromanipulator (cathode). The single secondary particle was then cycled in galvanostatic mode with current density calculated using an estimation of the particle volume (Figure 2e). The FIB/SEM is used to observe the internal evolution of the NMC secondary particle induced by the deliathiation process, either the particle is etched at the end of the cycles, or the particle is first half etched and then a thin slice is milled after each cycle (Figure 2f).

However, in this FIB/SEM micro-battery setup, some undesirable side effects should be considered during the experiment, as illustrated in Figure S2. Depending on the distance from the NMC secondary particle to the Li metal electrode, lithium dendrites can be formed from secondary particle surface and then touch the Li metal or holder inducing a short circuit (Figure S2a). The NMC secondary particle should be placed as close as possible to the Li electrode to optimise cycling, but not too close to avoid a short-circuiting dendrite formation. Moreover, it is worth noticing that the SEM electron beam has an incidence on the voltage level in the setup as shown in Figure S2b, in which the recharge of the battery is indirectly activated. Therefore, all electrochemical cycles were carried out with the electron beam off avoiding any disturbance. Moreover, Imaging and milling should then be monitored carefully to ensure that no extra damage or too strong changes are induced in the secondary NMC particle. Finally, if the cycling current is too strong dendrite loops attached to the particle surface can be formed in the electrolyte (Figure S2c). Due to high cycling voltage and the fact that only one particle is cycled, the electrolyte is likely to be degraded and a thick electrolyte layer is witnessed surrounding the particle at the end of cycling. Such a thick SEI layer is likely to build up with every cycle the secondary particle goes through and could significantly hinder its cycling efficiency when it becomes too thick. However, in this type of experiment with a single secondary particle, this SEI layer is also believed to help provide a mechanical coherency against crack propagation during the 3D FIB/SEM acquisition by keeping the primary particles from detaching and falling. Such a mechanical constraint could also happen in densely packed full electrodes.

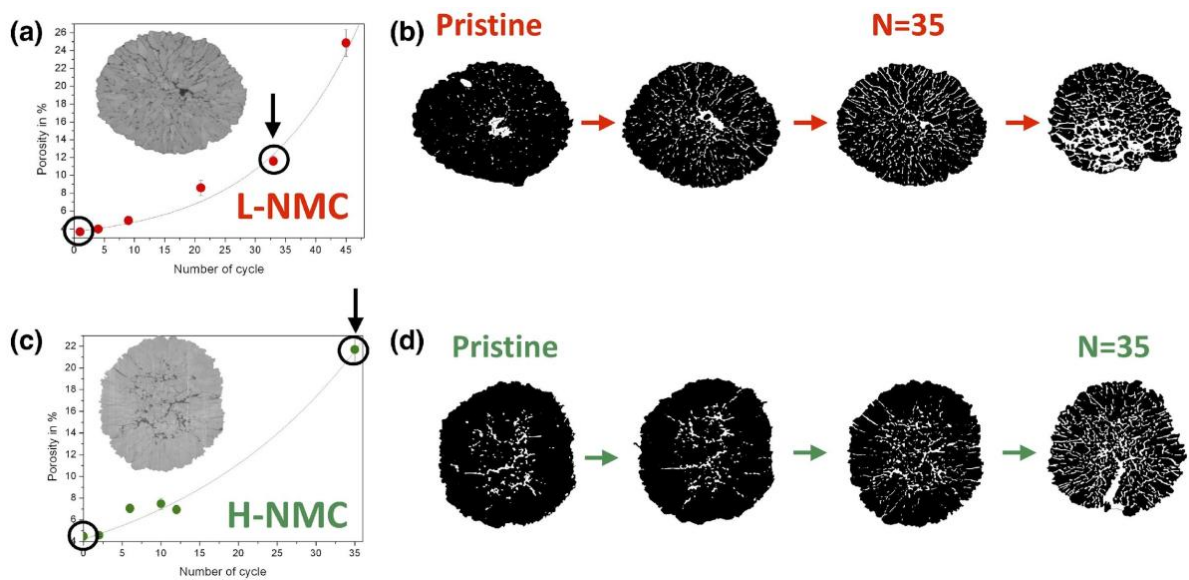

**Figure 3:** Evolution of the internal cracking structure within the NMC particles with the number of cycles. (a) and (c) show the porosity evolution for respectively L-NMC and H-NMC particles while (b) and (d) are, also respectively, displaying the crack structure evolution during particle ageing (white is porosity and black is NMC). Arrows and circles are indicating the pristine and 35[th] cycle for both particle types.

Several particles were cycled and monitored *in situ* for both L-NMC and H-NMC secondary particles and the porosity (cracks) evolution was examined as a function of the cycle number as exhibited in Figure 3. Experimental points were fitted using exponential decay function. This experiment was repeated between 5 to 10 times for each secondary particle type in order to check the uniformity of internal particle morphology. This morphological study, based on FIB thin slicing after each cycle, shows that the cracks and porosities evolving faster in the H-NMC secondary particles with around 22 $\%_v$ porosity at cycle 35 against 12 $\%_v$ for L-NMC. L-NMC secondary particles reach a similar porosity around cycles 42 to 45. Figures 3b and 3d illustrate it qualitatively with a view of porosity evolution and crack propagation within a slice of the secondary particles at different times. Crack propagation appears to be more stable in L-NMC than in H-NMC in which it leads faster to particle damage. This suggests that the initial porosity morphology can influence the crack propagation and formation within the NMC secondary particles. The NMC primary grain orientations and arrangements also play a crucial role as the cracks seem to only appear and propagate along the grain boundaries between primary particles. This is related to the synthesis process, both nucleation and growth parts of it, that was different between L-NMC and H-NMC secondary (and primary) particles along with their core shell structures.

An unexpected feature noticed during cycling is the increase in capacity that occurs during the first cycles (Figure 4b and c). Galvanostatic results (Figure 4b) showed that the capacity end-of-discharge increased with the number of cycles. The possible cause of this phenomenon is a reduction of polarisation related to charge transfers and solid-diffusion processes as there is no limitation from mass transport. Indeed, only one single secondary particle is considered here, no tortuosity should be considered for mass transport. To get a better understanding of this unexpected capacity rising, impedance spectroscopy (EIS) is used to follow transfer properties as a function of the cycle. The EIS results Figure 4c show a reduction of the high frequency semicircle diameter, which is attributed to charge transfer resistance reduction. An active surface area increase (NMC/electrolyte interface) through internal cracking could be an explanation as seen in Figure 3 and 4a. It is believed that this initial network of cracks allows, through creation and enlargement, deeper and faster access to the interior of the secondary NMC particles. This morphological change results in an increase in the NMC/electrolyte interface without excessively damaging the connections between the primary NMC grains composing the secondary particle. Additionally, more reaction sites on the surface of secondary particles may also affect solid diffusion as diffusion pathways to vacant pathways decrease[34,35]. The methodology used to fit EIS semi-circles is explained in detail in SI.

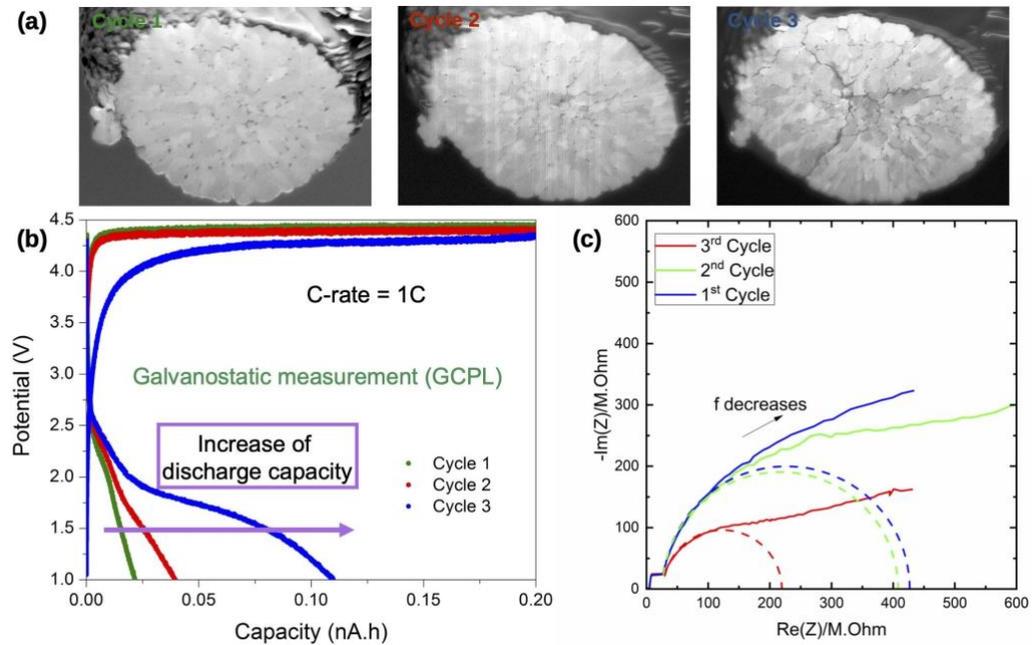

**Figure 4:** Evolutions of the internal crack structure and electrochemical properties during the first three cycles. The crack apparition and widening (a) appears to be related to the increase in discharge capacity (b) and the ionic resistance decrease (c) measured via impedance.

Such behaviour can provide a better access path for the Li-ion in the NMC active material, especially to the core grains, without separation of the primary grains to the main secondary particle. Indeed, this separation isolates the primary grains and is therefore detrimental to the cohesion of the secondary particle affecting the inner conductivity and the ability to store electrochemical energy. However, this degradation by splitting of secondary particle fractionation occurs in later cycles inducing expected capacity fading.

Figure 5 shows galvanostatic curves after many numbers of delithiation/lithiation cycles of NMC active materials with its solid-state transition from 2.0 to 4.8 V. Figure 5a illustrates potential (V) as a function of capacity (nA.h) curves for several cycles of a single secondary L-NMC particle. The single particle discharge behaviour can be separated into two regions. The first one gathers the very first cycles (~ 10) that show a highest capacity end-of-discharge. Then, the second region relates to the following cycles where the capacity is reduced along with an increase in polarisation. The better performances in the first regions observed here is consistent with an increase in active surface area as discussed above (Figure 3 and 4). As for the last cycles, the reduction in capacity end-of-discharge could be explained by a connection loss between different parts of the secondary NMC particles that are separated by the secondary particle cracking. This also leads to higher polarisation as the current per active surface area unit increases. In fact, maintaining the same discharge current but with lower connected NMC amount is equivalent to cycle with a higher C-rate and heavy secondary particle cracking can be witnessed along cycling in Figure 3. Such cracking behaviour is believed to be exacerbated here by the cycling of one single secondary particle. However, it should still be representative of an entire positive electrode with modifications condensed in a smaller time range and with a kind of optimization of the electronic conductivity percolation network[8].

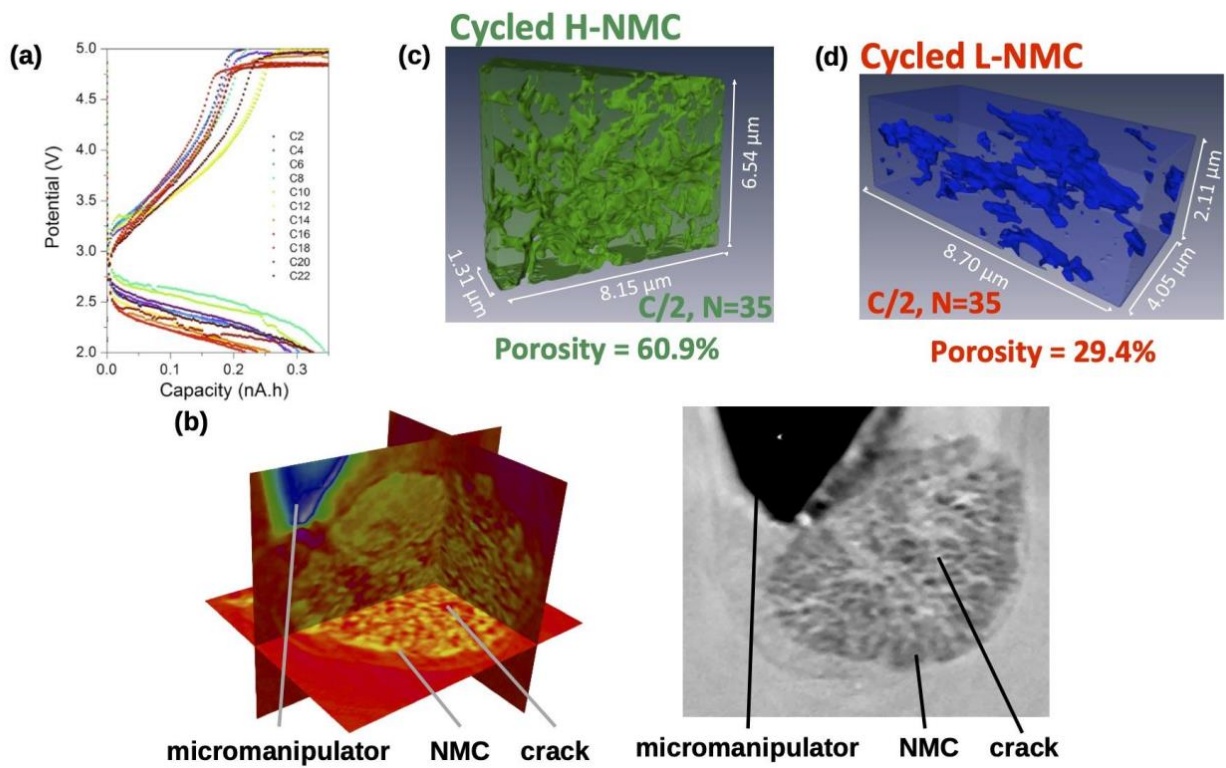

**Figure 5:** Cycling effects in the NMC particles. a) is gathering the cycling curves for the different cycles C in L-NMC. b) is a 3D XR tomography view of a cycled single secondary L-NMC particle after 35 cycles. Some cracks within the secondary particle can be witnessed. c) and d) show the internal porosity after cycling at C/2 in the regions of interest similar as those defined in Figure 1 for respectively H-NMC and L-NMC.

Nano-CT TXM[36] data acquired at the APS synchrotron (ID32 beamline) shows qualitatively (Figure 5b) the presence of a crack network in a cycled (35 cycles) L-NMC secondary particle without any incidence of electron beams during acquisition. Figures 5c and 5d are bringing a closer look at the porosity phase after 35 cycles at C/2 C-rate for, respectively, H-NMC and L-NMC in a wide region of interest within the secondary particles. These microstructures were acquired with the FIB/SEM and evidenced a more widely spread and intra-connected crack network (and a much higher porosity, 61 %$_v$ against 30 %$_v$) for H-NMC, compared to L-NMC, in similar conditions. Such a network then indicates a higher damage rate for the H-NMC secondary particles when compared to L-NMC ones which may arise from the differences in the initial porosity shape. Therefore, our study highlights that the initial internal porosity leads to different scenarios of 3D morphology evolution and subsequent cracking and splitting of secondary particles.

In conclusion, we present here an approach using *in situ* electrochemical FIB/SEM setup that allows single secondary particle cycling along with the investigation of its morphological changes during electrochemical cycling. It was applied to two types of active material secondary particles for Li-ion batteries: high and low gradient NMC. This methodology allowed us to capture, and explain, an enhancement in the cell electrochemical properties in the first cycles. The apparition and widening of an initial crack structure is, at that early stage, easing the ion flow toward the secondary particle core without damaging too much the grain boundaries between primary NMC grains as it is witnessed for later cycles and causing the overall final capacity to fade. The porosity and crack network evolution were also investigated in-situ for the same secondary particles. It is evident that the difference in initial porosity shape between low gradient and high gradient NMC secondary particles plays a role in the secondary particle damage over cycles. The bulkier porosity in the low gradient NMC secondary particles appears to create a crack network that is less detrimental to the particle structure than the spider web like shape in the high gradient ones. Porosity also increases more slowly in low gradient particles. This is also confirmed by the investigation of secondary particles cycled ex-situ with FIB/SEM and nano-XCT performed after full cycling. Therefore, the engineering of secondary particle morphology and porosity should be further considered to mitigate crack damage to improve the performance of cathode electrodes.

Associated content:
**Supporting information (pdf)**
Author information:
**Corresponding Authors**
E-mail: dean.miller@tescan.com
E-mail: arnaud.demortiere@cnrs.fr
**Author Contributions**
D.M. and A.D. have conceived the investigation and built the *in situ* set-up. This work was supervised by D.M. and A.D. The datasets were acquired by A.D. . FIB/SEM tomography datasets were reconstructed analyzed by J.A., F.C., Z.L.S. and A.D. . T.-T.N. and M.B. analyzed electrochemical datasets. V.D.

performed the nano-XCT experiments. A.D. and F.C. wrote the manuscript. All authors participated in the discussion and revision of this paper and finally approved this work.

**Notes**

The authors declare no competing financial interest

**Data Availability Statement**

Research data are not shared.

Acknowledgments:

This work was performed, in part, at the Center for Nanoscale Materials, a U.S. Department of Energy Office of Science User Facility, under Contract No. DE-AC02-06CH11357. The authors would like to acknowledge French National Association for Research and Technology (ANRT) for partially supporting the funding of this research work.

**Supporting information**

Materials and method:

The NMC secondary particles used here are gradient based of two types: low gradient (L-NMC) and high gradient (H-NMC). This means their composition varies from the inner to the outer shell. L-NMC are composed of $Li(Ni_{0.68}Co_{0.18}Mn_{0.18})O_2$ on average with a core $Li(Ni_{0.8}Co_{0.1}Mn_{0.1})O_2$ and a $Li(Ni_{0.46}Co_{0.23}Mn_{0.31})O_2$ outer shell while H-NMC are based on $Li(Ni_{0.75}Co_{0.1}Mn_{0.15})O_2$ on average with a core $Li(Ni_{0.86}Co_{0.1}Mn_{0.04})O_2$ and a $Li(Ni_{0.7}Co_{0.1}Mn_{0.2})O_2$ outer shell.

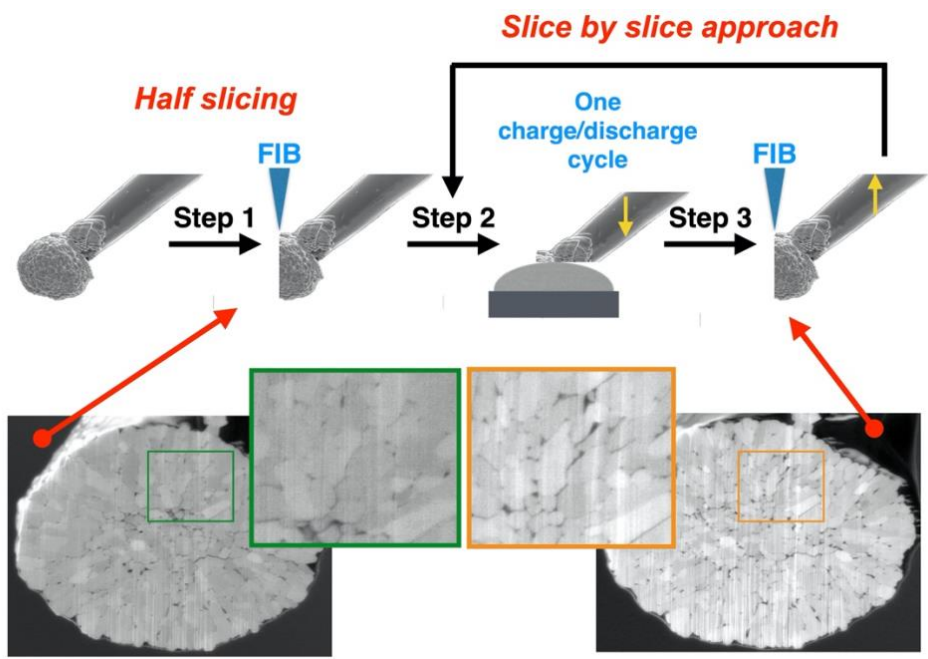

**Figure S1:** Schematic view of the in-situ approach to image structural change between each cycle in a FIB/SEM.

As described in the main article, the electrochemical cycling of single secondary particles is done inside a Zeiss SII Nvision40 FIB/SEM with an omniprob micromanipulator and consists in attaching a secondary particle to the FIB/SEM micromanipulator via a carbon bridge (which is electronically conductive). This carbon bridge layer was carefully placed to make sure it does not cover much of the electrochemically active areas in order to do not hinder the electrochemical cycling of the secondary particle. Lithium metal have also been previously deposited on the SEM holder and covered with low vapor pressure electrolyte (Ionic liquid), here a mixture of $P_{13}$ TFSI + 0.5 mol/kg LiTFSI[8]. The single secondary particle is then placed into the electrolyte layer making sure that it is neither too close nor too far from the Li metal. Li metal and the single NMC particle respectively acts as negative and positive electrodes. Compared to real use electrodes there is here no separator as the two electrodes are not supposed to be in contact. An ultra-low current instrument (biologic SP200) is used to perform the electrochemical measurements through a two-point contact being the micromanipulator and the SEM holder. After immersing into electrolyte and minimization of Li-metal/particle distance, the single particle of active materials is cycled in galvanostatic mode with a steady current depending on the desired C-rate.

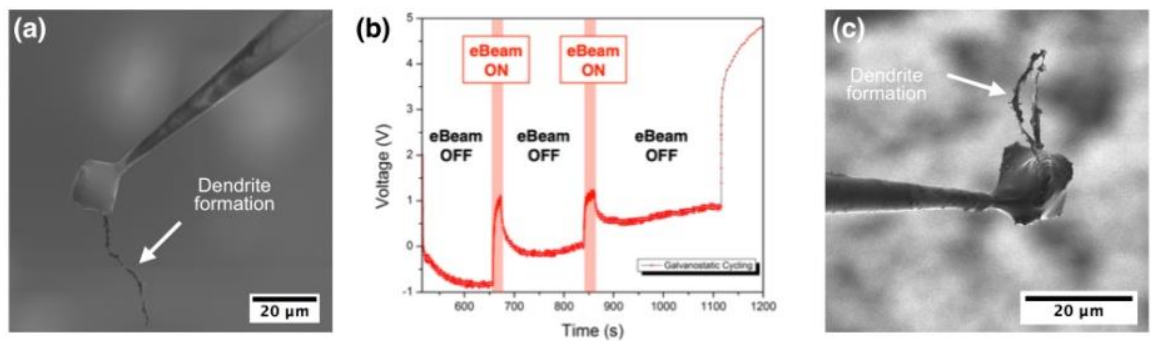

**Figure S2:** Side effect examples of phenomenon that occur during electrochemical reaction. a) and c) evidence of Li dendrite formations on the NMC particle surface and the SEM base holder/counter electrode. b) shows the electron beam effect on the cycling data.

The in-situ monitoring of the secondary particle morphology is allowed by performing an initial milling that cut the attached secondary particle in half to reveal its inner structure. Then the cut face is completely immersed in the electrolyte layer for cycling. At the end of each desired cycle, the particle is removed from the electrolyte a thin layer is milled using the FIB beam in order to reveal the microstructure before imaging it with SEM. This milling step is indeed necessary to remove the SEI/electrolyte coating the particle surface. The milled layer must remain thin enough to keep the captured base microstructure as close as possible to the previous one, typically here 19 nm (300 pA milling current) were milled at each imaging step (13.5 x 13.5 nm² pixel size). This methodology leads to a stack of 2D images which goal is to image the microstructural state of the same location within the secondary particle through time and cycles.

Some 3D microstructures were also acquired with FIB/SEM to image full secondary particles. This was done is the same FIB/SEM with a 13.5 x 13.5 nm² pixel size for SEM imaging and a 19 nm milling thickness (same parameters as for the in-situ experiments). X-Ray tomography data were acquired at the 32-ID APS synchrotron beamline at Argonne NL. The data set was acquired at 8 keV with 3 s exposure time/projection. X-ray objective lens = Fresnel zone plate (FZP) with 60 nm outermost zone width. The numerical aperture of the FZP was matched by the condenser. The condenser was a beam shaping condenser, i.e., a grating with 1.32 mm of diameter and 60 nm spacing for the outermost grating. Approximately 710 projections acquired along 1770 instead of the usual 1800. We had missing angles because of the substrate blocking the view. Pixel is ≈ 20 nm large but the true spatial resolution is 60 nm (you can see 60 nm features that are at least 2 voxels wide).

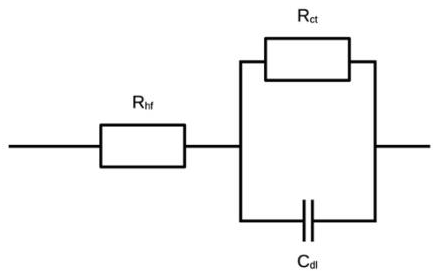

**Figure S3:** Equivalent electronic circuit used to fit the EIS data.

For the determination of charge transfer resistance, an electric circuit containing a resistance ($R_{hf}$) in series with a parallel resistance/capacitor setup ($R_{ct}$//$C_{dl}$) was employed to fit the high frequency feature (arc), *cf.* Figure S3. This circuit is a simple model for an electrode immersed in an electrolyte. The resistance $R_{hf}$ represents either the ionic or contact resistances that might arise during the measurement. A faradaic reaction is assumed to occur at the solid/liquid interface in parallel with the charging of the double layer. Therefore, the charge transfer resistance $R_{ct}$, associated with the faradaic reaction, is modeled in parallel to $C_{dl}$. As a result, the semi-circle diameter gives the value of the charge transfer resistance.